\documentclass[twocolumn,prl,aps,superscriptaddress,epsfig]{revtex4}
\usepackage{epsfig}
\usepackage{dcolumn}
\begin{document}
\title{Disorder driven collapse of the mobility gap and transition to an 
insulator in fractional quantum Hall effect}
\author{D. N. Sheng}
\affiliation{Department of Physics and Astronomy, California State University,
Northridge, California 91330}
\author{Xin Wan}
\affiliation{National High Magnetic Field Laboratory and Department of Physics,
Florida State University, Tallahasse, Florida 32306}
\author{E. H. Rezayi}
\affiliation{Physics Department California State University Los Angeles, 
Los Angeles, California 90032}
\author{Kun Yang}
\affiliation{National High Magnetic Field Laboratory and Department of Physics,
Florida State University, Tallahasse, Florida 32306}
\author{R. N. Bhatt}
\affiliation{Department of Electrical Engineering, Princeton University, 
Princeton, New Jersey 08544}
\author{F. D. M. Haldane}
\affiliation{Department of Physics, Jadwin Hall, Princeton University, 
Princeton, New Jersey 08544}
\date{\today}
\def\PRL#1{{\sl Phys. Rev. Lett. }{\bf #1}}
\def\PRB#1{{\sl Phys. Rev.  B }{\bf #1}}
\begin{abstract}
We study the $\nu=1/3$ quantum Hall state in the presence of random disorder.
We calculate the topologically invariant Chern number, which is the only 
quantity known at present to unambiguously distinguish between insulating and current 
carrying states in an interacting system. The mobility gap can be 
determined numerically this way, which is found to agree with experimental 
value semiquantitatively.
As the disorder strength increases towards a critical value, both the mobility 
gap and plateau width narrow continuously and
ultimately collapse leading to an insulating phase.
\end{abstract}
\maketitle
Two-dimensional electron systems in a perpendicular magnetic field
have been the focus of both theoretical and experimental attention for the 
past two decades. Such  systems, if 
made sufficiently pure and taken to low enough temperatures, exhibit 
primarily  the fractional quantum Hall effect\cite{qhe1,qhe2,qhe3} (FQHE).
While the pure systems, especially for the strongest FQHE, are fairly well
understood, there is essentially no quantitative 
understanding of the role of random disorder.  The importance of disorder, 
however, is underscored by the hallmark FQHE plateaus, which can not occur 
in a translationally invariant system. On the other hand, when the disorder 
strength becomes comparable to the strength of interactions between electrons,
the FQHE will eventually be destroyed.  In this paper, we report
on finite-size numerical studies of the effects of random disorder on the 1/3 FQHE.

The usual numerically calculated quantities such as the energy spectrum, 
wavefunctions, and the density-density correlation functions, while useful in 
studies of isolated impurities\cite{das,rez}, provide little or no 
understanding of transport properties. A more appropriate approach is to 
obtain the quantum Hall conductance by calculating topologically invariant 
Chern integers\cite{chern1,chern2,chern3} in systems with periodic
boundary conditions (or torus geometry), in the presence of 
random disorder. Physically the Chern integers are the boundary condition 
averaged Hall conductance of the system in units of $e^2/h$.
In the case of the integer quantum Hall effect (IQHE), 
the Chern numbers for the ground states have 
fixed non-zero integer values, while a ``metallic" or critical state that
separates two neighboring IQHE states gives an intrinsically 
fluctuating Chern number\cite{chern3}.
(It takes on different integer values in response to slight changes 
of the external parameters, such as the disorder configuration). Such a 
quantum phase transition has only been studied for several 
{\em noninteracting} models\cite{chern3}, in which only the Chern number 
for single particle states needs to be calculated. 

The situation is very different in the case of FQHE.
Due to the topological and
many-body nature of the problem, in the FQHE system at filling 
factor $\nu=nh/eB=p/q$ ($B$ is the magnetic field and $n$ the areal density), 
there exists a manifold of $q$ nearly degenerate low energy states on the
torus, whose 
energy differences disappear in the thermodynamic limit 
(Ref. \onlinecite{wenniu} and see below).
The quantization of the Hall conductance can not be tied, in a physically
meaningful way, to any particular one of the $q$ ground states. 
For example, in the absence of disorder (in which case the degeneracy is 
exact), when an external flux quanta is inserted adiabatically\cite{bob,chern1,tao} 
in a region inaccessible to 
the electrons, the states within a given manifold evolve into each other.
In the presence of disorder, the Chern number of the individual states 
fluctuates while the sum of the Chern number of all the states 
turns out to be $p$ and robust.
As a consequence, we may regard  the total Chern number $p$ to be shared
by the $q$ degenerate states, which results in fractionally
quantized Hall conductance $\sigma_H=pe^2/qh$ 
for the system. 

Based on this picture, we have developed a numerical method for studying
the topological Chern number of the ground state and low energy
excited states for an {\em interacting} system. 
We will show that the presence of disorder leads to several interesting and
important results for 1/3 FQHE:
(i) A weak random
disorder lifts the degeneracy of the ground state for a finite
number of electrons. However,
the level spacings between the lowest three states decrease monotonically
with the increase of electron number, indicating  the recovery of the
degeneracy  in the thermodynamic limit. 
(ii) The mobility gap, which separates the higher energy extended excitations 
from the low energy FQHE states, can be determined from the distribution of the 
Chern number of the many-body states. This  is the only way  
to unambiguously distinguish between insulating and current 
carrying states in an interacting system. 
(iii) In general the mobility gap, determined this way, will be different from the 
spectrum gap (which separates the lowest three states from other higher states).
It is the mobility gap that should be compared 
with the experimentally obtained activation energies.
(iv) There exists a 
critical disorder strength
$W=W_c\approx 0.2 e^2/\epsilon$ for Gaussian white noise potential, 
which marks a transition from the FQHE to insulator.
(v) The physics of the destruction of the FQHE can be described as
the continuous collapse of the mobility gap;
the closing of the mobility gap
and quantization of the Hall conductance occur at the same time.

We consider a two-dimensional interacting electron system in an 
$L_1\times L_2$ square cell
with twisted boundary conditions: 
$T({\bf L}_j)\Phi({\bf r})=e^{i\theta_j}\Phi ({\bf r})$, 
where $T({\bf L}_j)$ is the magnetic translation
operator and $j=1,2$ represents the $x$ and $y$ directions,
respectively.  
Calculating the Hall conductance $\sigma_H$ directly from
a Kubo formula would require a knowledge of all the many body eigenstates 
for a fixed number of electrons.
This proves impractical for systems with 3 electrons or larger.
As first realized by Thouless\cite{chern1} and co-workers, a topological 
property of the 
wavefunction\cite{bob}, known as the first Chern number,
can be used to calculate the boundary condition averaged 
$\sigma_H$. 
The importance of Chern numbers, however, goes beyond obtaining $\sigma_H$.  It appears to be 
the only quantity that distinguishes 
between insulating and current carrying states\cite{chern2,note} in an 
interacting system. 
Because we are dealing with many-body wavefunctions,
other simpler numerical methods (such as inverse participation ratio and 
Thouless numbers) used for 
determining the localization of the single-particle wavefunctions 
have no obvious extension here. 
After a unitary transformation $\Psi_k=exp[-i \sum _{i=1}^{N_e}(\frac
 {\theta_1}{L_1} x_i +\frac {\theta_2}{L_2}y_i)]\Phi_k $,
the boundary-phase averaged Hall conductance 
for the $k$-th many-body eigenstate is $\sigma_H^k= C(k)e^2/h$, 
with
\begin{eqnarray}
\label{integral}
C(k) ={i\over 4\pi}\oint d\theta_j
\{\langle { \psi_k |{\partial \psi_k 
 \over
\partial \theta_j}\rangle -  
\langle {\partial \psi_k \over \partial \theta_j}| 
\psi_k} 
\rangle\},
\end{eqnarray}
where the closed path integral is along the boundary of a unit cell
$0 \leq \theta_1, \theta_2 \leq 2\pi$ and summation over $j$ is implied.
$C(k)$  is exactly the Berry phase (in units of $2\pi$) accumulated for
such a state when the boundary phase evolves along the closed path.
To  determine the Chern number uniquely\cite{chern3},
we separate the   boundary phase space
into approximately  $36-100$ mesh
points and get the sum of the Berry phases from each mesh.
In cases where there are near-level-crossings, the integration contour has to be
chosen reasonably close to these points.  This determines the size of the mesh, 
at least locally.
For the mesh sizes we chose, we found the Chern numbers had converged
and did  not change by further reducing the size of the mesh.

In the presence of a strong magnetic field, one can project
the Hamiltonian onto the  partially-filled, lowest  Landau level.
The projected Hamiltonian in the presence of both Coulomb interaction
and disorder\cite{sheng} can be written as:
\begin{eqnarray} 
H&=&\sum _{i<j} \sum _{ \bf {q}}  e^{-q^2/2 }  V(q)
e^{i {\bf q} \cdot ({\bf R}_i -{\bf R}_j)} \nonumber \\ 
&+&\sum _{i} \sum _{ \bf q} e^{-q^2/4} U_{\bf q} 
e^{i{\bf q }\cdot {\bf R}_i},
\end{eqnarray}
where ${\bf R}_i$ is the guiding center coordinate of the $i$-th
electron, $V(q)=2\pi e^2/\epsilon q$ is the
Coulomb potential, and $U_{\mathbf q}$ is the impurity potential
with the wave vector ${\bf q}$.
We set the magnetic length $\ell=1$ and $e^2/\epsilon\ell=1$ for convenience.
The Gaussian white noise potential we use is generated according to the 
correlation relation
in $q$-space 
$\langle U_{\bf q}U_{{\bf q}'}\rangle=(W^2/A) \delta_{{\bf q},-{\bf q'}}$,
which corresponds to $\langle U({\bf r})U({\bf r'})\rangle
=W^2 \delta (\bf {r-r'})$
in real space, where $W$ is the strength of the disorder and $A$ is the area of
the system.
We consider the case of $\nu =N_e/N_\phi
=1/3$, where $N_e$ and $N_\phi$ are the number of electrons and 
flux quanta.
We obtain the exact low energy eigenvalues and eigenstates 
using the Lanczos method 
for systems up to  $N_e=8$ electrons in 24 flux quanta, spanning a Hilbert 
space of size $N_{\rm basis}=735,471$. 
We then calculate the Chern numbers using Eq. (1).

Independent of boundary conditions, we find that, for weak disorder,
low energy states are separated into groups. The lowest group has 
three closely spaced states. This property comes from the
three-fold center of mass degeneracy of the pure system at $\nu=1/3$
and is shared by the entire spectrum.
For finite sizes, there is a  finite level spacing between
the lowest three states, which is much smaller than the energy difference
between the 3rd and the 4th state. The latter is the low energy
spectrum gap denoted as $E_{s}$.  At $W=0.06$, we have 
$E_s=0.04\pm 0.002$ (averaged over 100-2000 disorder configurations), which
is size independent for $N_e=5-8$.   
It remains finite for weak W until  $W$ is further increased
to $W\approx 0.15$, at which point $E_s$ becomes too small and
its value in the thermodynamic limit cannot be extrapolated from
the sizes accessible to our approach.
\begin{figure}
\begin{center}
\vspace*{1.5cm}
\epsfxsize=8.5truecm \epsffile{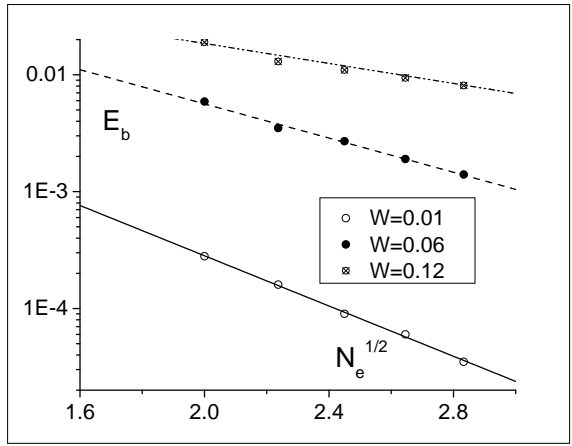}
\vspace*{-2.5cm}
\caption{
A semilog plot of the bandwidth $E_b$ of the first level that consists of three closely spaced lowest energy states
as a function of the square root of the electron number $N_e$,
at three relatively weak  disorder strengths
$W=0.01$, $W=0.06$ and   $W=0.12$. }
\label{fig:fig1}
\end{center}
\end{figure}

On the other hand, the level spacings between the lowest three states depend 
strongly on $N_e$. They are expected to 
vanish exponentially as the linear dimension
of the system\cite{wenniu}. This can be seen clearly in
Fig. 1, where we show a semilog plot of the ``bandwidth'' $E_b$ vs $\sqrt{N_e}$ 
for three different disorder strengths.
It is apparent that $E_b$ drops to zero for large $N_e$,
a direct consequence of the topological order 
in the ground state.  
This is in contrast to the usual effect of disorder in the
IQHE, where the degeneracy of all the states in a Landau level,
not being a topological property,
will be lifted by the perturbation of arbitrarily weak disorder.
\begin{figure}
\begin{center}
\vspace*{0.2cm}
\hspace*{5mm}
\epsfxsize=8.5truecm \epsffile{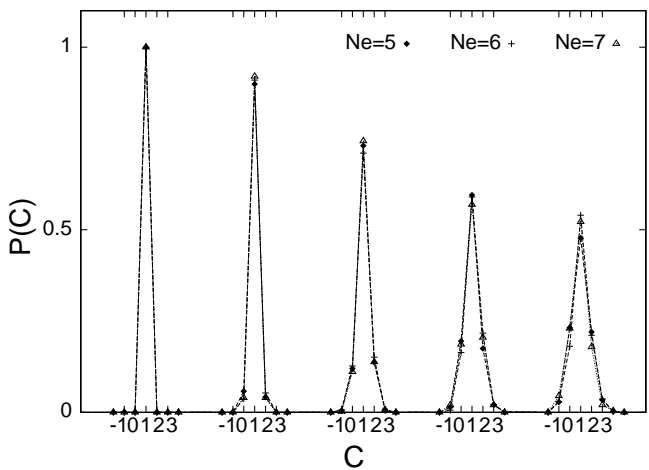}
\vspace*{-0.6cm}
\caption{
The probability distribution $P(C)$ of total Chern number $C$,
for the lowest five  groups of states.
The energy of the states increases towards the right.}
\label{fig:fig2}
\end{center}
\end{figure}

In the presence of weak disorder ($W<0.12$), we find that the total Chern number
carried by the lowest level is always 1, for 
$N_e=3-8$ and thousands of disorder configurations. 
Note that the vanishing $E_b$ 
and the existence of a finite spectrum gap $E_s$, at relatively
weak $W$, is a direct manifestation of the
$\nu =1/3 $ FQHE; since the
lowest three states become degenerate in the thermodynamic limit,
the Hall conductance carried by the ground state
is $1/3$ of $e^2/h$.
However, this is not unique to the ground state and we find that even the low
lying  excited  states above the spectrum gap have similar
properties. 
Namely, each three states carry 
total Chern number $C=1$.  At $W=0.01$, the lowest two groups
have $C=1$ for all the disorder configurations we have sampled.
In Fig. 2 we plot for the first five groups, $P(C)$, for $W=0.06$,   
vs $C$. $P(C)$ is the probability that the total group Chern number is $C$.
We have averaged over 2000, 1000 and 500 disorder configurations for 
$N_e=5, 6 $ and 7, respectively. 
As seen in Fig. 2, $P(1) =1$ for the first group ($N_g=1$),  
and $P(C)=0$ for $C\neq 1$. This means
all disorder configurations have $C=1$, 
which  corresponds to the 1/3 FQHE because each state
carries a definite  average Hall conductance of $e^2/3h$.
For $N_g=2$,
both $P(0)$ and $P(2)$ are nonzero, indicating that 
a small number of configurations have $C=0$ or 2.
As a result, $ P(1)$  is reduced to $0.9,\ 0.91$, and $0.92$ for $N_e=5,\ 6$,
and 7. The increase of $P(1)$ with $N_e$
indicates that $P(1)$  may recover to 1 at large
$N_e$. 
For the $N_g=3$ case,
$P(1)$ is significantly reduced to about 0.7; it behaves 
non-monotonically as a function of $N_e$.  This  results from the coexistence
of three different Chern numbers in the thermodynamic limit,
which characterizes the delocalization of quasi-particle excitations.
Namely, these excitations carry nonzero Chern numbers which
are extended in real space. 
For $N_g>3$,  $P(1)$ is further reduced and seems to saturate at a value
near 0.5. 

We may regard the fluctuation of the Chern number to be
an indication of the degree of delocalization. 
In analogy to the physics of noninteracting systems\cite{chern3},
we define 
$P_{ext}=1-P(1)$ to be the likelihood of the breakdown of the quantization
of the Hall conductance and thus a measure of the delocalization of the
charged  excitations\cite{note}. 
In the FQHE plateau regime,  $P_{ext}$ goes to zero as a result of 
the localization  or nondissipative nature of the state.
Here nonzero $P_{ext}$  occurs as we go to higher energy states
 ($N_g \geq 3$).
The energy that separates these two kinds of states is called 
the mobility edge, where $P_{ext}$ has a large increase,
which probably becomes a finite jump in the thermodynamic limit.
For $W=0.06$, we find that, from $N_g=2$ to 3, $P_{ext}$
has the largest increase, which puts the $N_g=3$ group at the mobility
edge. 
Measuring the energy at the mobility edge relative to the 
energy of the lowest level,  we get the mobility gap $E_m$ for $N_e=4-8$,
which is $N_e$-dependent.
For $W\ge 0.06$, we 
determine the mobility gap by extrapolating 
the finite size data to the thermodynamic limit. 
For weaker disorder, the sizes that can be treated are 
not sufficient to produce a meaningful extrapolation to $N\rightarrow \infty$. 
\begin{figure}[h!]
\epsfxsize=8.0truecm \epsffile{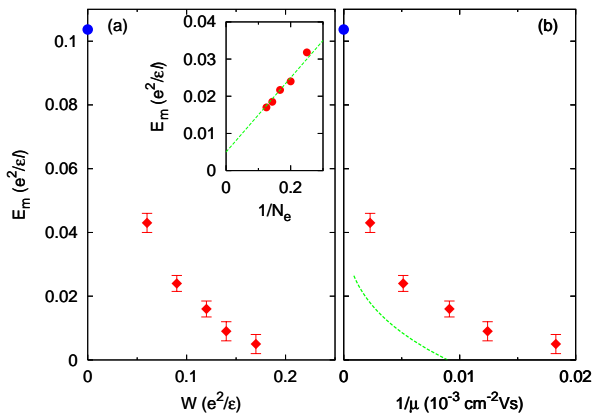}
\caption{
\label{mobilitygap}
(a) The extrapolated mobility gap $E_m$ as a function of $W$. 
Inset shows $E_m$ at $W = 0.17$ for $N_e =$ 4-8 electrons. 
The dashed line indicates $E_m$ can be extrapolated to
0.005 at $1/N_e \rightarrow 0$.
The dot at $W = 0$ is the creation energy for a quasiparticle-quasihole
pair at infinite separation, extrapolated from pure systems with up to 
$N_e = 10$.   
(b) Dependence of $E_m$ on mobility $\mu$. 
The dashed line is converted from a fit to experimental
data (taken from Ref.~\onlinecite{boebinger87}). 
Here, we use an empirical mobility-density relation as well as 
a mobility-disorder relation in the Born approximation 
(see text for detail). 
}
\end{figure}

The extrapolated $E_m$ vs. $W$ is shown in Fig.~\ref{mobilitygap}a.
In the inset we plot $E_m$ vs. $1/N_e$ for $W=0.17$,
which can be best fit to $E_m = 0.10/N_e+0.005$;
thus we obtain $E_m=0.005\pm 0.003$ in the thermodynamic limit.  
We see that at  $W\approx 0.17$ such a gap is strongly reduced,
consistent with the drop of the spectrum gap for similar $W$,
signaling the FQHE is on the verge of being destroyed by disorder.

The energy gap $\Delta$ in the excitation spectrum of the correlated many-body
ground state can be extracted experimentally from the temperature
dependence of the magnetoresistivity, $\rho_{xx} \propto \exp (-\Delta /
2 k_B T)$, where $\Delta / 2$ is the activation energy and $k_B$ the
Boltzmann's constant~\cite{boebinger87,willett88}. 
Boebinger {\it et al.}~\cite{boebinger87} 
systematically studied the activation energy for
$\nu = 1/3$, 2/3, 4/3, and 5/3 and its dependence on sample
mobility $\mu$ (an indication of disorder) in GaAs-Al$_x$Ga$_{1-x}$As. 
For a class of high-mobility samples, they found that $\Delta(\mu) \simeq
C_0(\mu) (e^2/\epsilon l) - \Gamma(\mu)$, 
consistent with a simple phenomenological model~\cite{chang83}
that assumes a disorder-broadened excitation energy level with
half-width $\Gamma(\mu)$.
Figure~\ref{mobilitygap}b compares this empirical formula of $\Delta(\mu)$
with fitting parameters ($C_0 = 0.049$ and 
$\Gamma = 6$K~\cite{boebinger87}) for high-mobility
samples with the mobility gap we obtained in our calculation. 
To do this, we assume that both the (zero field) mobility and (high field)
mobility gap are dominated by short-range
scatterers (appropriate for high-mobility samples) and, in the Born 
approximation,
$\mu = e \hbar^3 / (m^{*2} W^2)$.
We also use an empirical relation $\mu \propto n^{1.5}$
between $\mu$ and electron density $n$
(both for zero gate bias) extracted from Fig.~1 of
Ref.~\onlinecite{boebinger87}.  
Here, we do not include the effects of layer thickness and Landau level
mixing which, nevertheless, exist in experimental samples 
and are known to reduce the gap by as much as 
a factor of 2~\cite{yoshioka86,willett88} (for zero or very weak disorder).

\begin{figure}
\begin{center}
\vspace*{1.5cm}
\epsfxsize=8.5truecm \epsffile{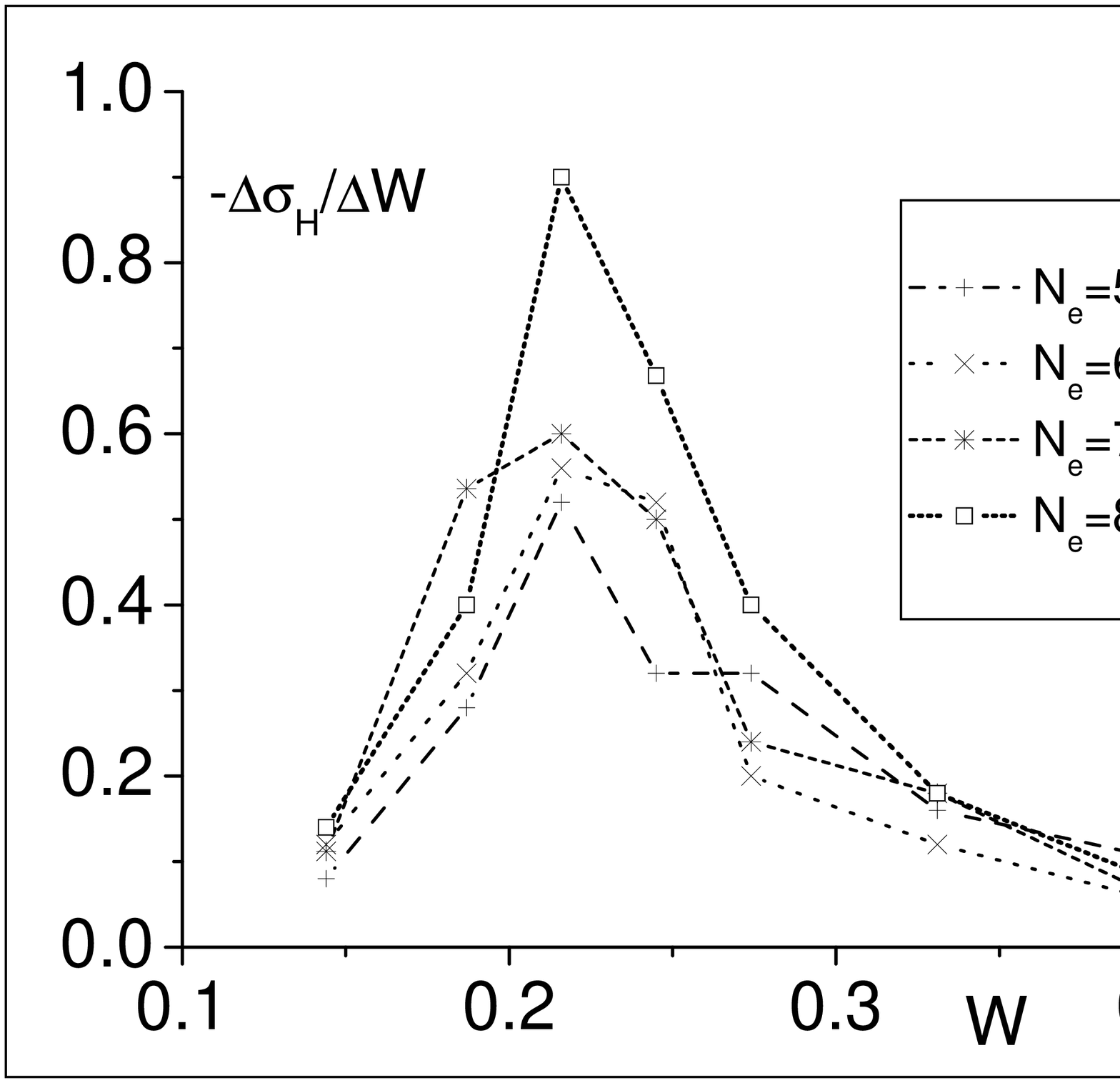}
\vspace{-2.5cm}
\caption{
The relative decrease  of the Hall conductance 
of the lowest level 
over the change of disorder strength 
$\Delta W$  is plotted 
as a function of disorder strength $W$ for $N_e=5,6,7$ and $8$.
}
\end{center}
\end{figure}

As we further increase W, the
FQHE becomes unstable. This can be discerned 
by following the evolution of the  Chern number 
of the lowest level
and  $\sigma_H$ averaged over  
the lowest three states. 
For example, for $N_e=6$,  at $W \leq 0.14$, we have $\sigma_H=e^2/3h$;
it drops to $0.924e^2/3h$ at  $W=0.17$. 
At larger $W$,
we find  a  very strong enhancement
of the fluctuation in the Chern number and, correspondingly, 
a rapid reduction of $\sigma_H$.
Similar results
are obtained  for $N_e=8$. 
As shown in Fig. 4,  $-\Delta \sigma_H /\Delta W$ has its largest value near 
$W_c=0.22\pm 0.025$ for all $N_e=5-8$, 
which determines the critical disorder for the $\nu=1/3$ state plateau
to insulator transition.

To illustrate the nature of the phase transition,              
we change the filling to be slightly off 1/3.
At weak $W$, the $\nu=1/3$ FQHE plateau
has a finite width due to the nonzero mobility gap that  survives
to fillings slightly below and above 1/3.
But at  $W=0.17$, which is slightly below the critical
$W_c$, we find the mobility gap $E_m$ at $N_e=N_s/3+1$ is already
reduced to zero. This suggests that the plateau for disorder close to $W_c$
is still pinned at filling $\nu=1/3$, but with zero width. Therefore, 
the disappearance of the FQHE
at $W_c$ is caused by the collapse of the mobility gap.

We would like to acknowledge helpful discussions with L. Balents, Mathew Fisher, X. G. 
Wen,  and Z. Y. Weng.
This work is supported by ACS-PRF \# 36965-AC5 and Research Corporation Award
CC5643 (DNS), by DOE under contract DE-FG03-02ER45981 (EHR),
by NSF grants DMR-9971541 and DMR-0225698 (XW and KY), 
DMR-00116566 (DNS), DMR-0213706 (RNB andFDMH),
the State of Florida (XW), and the A. P. Sloan Foundation (KY).
We acknowledge NPACI and FSU CSTI supercomputer support.
 
\end{document}